\documentclass[aip,jcp,reprint]{revtex4-1}
\usepackage{graphicx}
\usepackage{dcolumn}
\usepackage{bm}
\usepackage{amsmath}
\usepackage{amssymb}
\usepackage{float}
\usepackage{epsfig}
\usepackage{color}

\usepackage{url}

\usepackage{epstopdf}

\begin{document}

\title{A Diagrammatic Kinetic Theory of Density Fluctuations in Simple Liquids in the Overdamped Limit. II. The One-Loop Approximation}
\author{Kevin R. Pilkiewicz}\email{pilkman@gmail.com}
\author{Hans C. Andersen}\email{hca@stanford.edu}
\affiliation{Department of Chemistry, Stanford University, Stanford, California 94305}

\begin{abstract}
A diagrammatic kinetic theory of density fluctuations in simple dense liquids at long times, described in the preceding paper, is applied to a high density Lennard-Jones liquid to calculate various equilibrium time correlation functions.  The calculation starts from the general theory and makes two approximations.  1.  The general diagrammatic expression for an irreducible memory kernel is approximated using a one-loop approximation.  2.  The generalized Enskog projected propagator, which is required for the calculation, is approximated using a simple kinetic model for the hard sphere memory function.  The coherent intermediate scattering function (CISF), the longitudinal current correlation function (LCCF), the transverse current correlation function (TCCF), the incoherent intermediate scattering function (IISF), and the incoherent longitudinal current correlation function (ILCCF) are calculated and compared with simulation results for the Lennard-Jones liquid at high density.  The approximate theoretical results are in good agreement with the simulation data for the  IISF for all wave vectors studied and for the CISF and LCCF for large wave vector.  The approximate results are in poor agreement with the simulation data for the CISF,  LCCF, and TCCF for small wave vectors because these functions are strongly affected by hydrodynamic fluctuations at small wave vector that are not well described by the simple kinetic model used.  The possible implications of this approach for the study of liquids is discussed.

\end{abstract}

\maketitle

\section{Introduction}

In the previous paper in this series,\cite{PIL14A} which we shall refer to as paper I, we started from an exact graphical kinetic theory for the correlation function of phase space density fluctuations in a dense atomic fluid and demonstrated how making a certain set of well-defined assumptions about the short time scale dynamics of the system, in conjunction with a well specified long time limit, leads to a much simpler theory for longer time scales. We refer to this theory as the overdamped theory. In this paper we test the overdamped theory by comparing its predictions with simulation results for a dense Lennard-Jones fluid at a variety of temperatures.
 
 The properties we calculate from the theory are five basic time correlation functions for an atomic liquid: the coherent intermediate scattering function (CISF) $\phi_\rho(q,t)$, the longitudinal current correlation function (LCCF) $\phi_{jl}(q,t)$, the transverse current correlation function (TCCF) $\phi_{jt}(q,t)$, the incoherent intermediate scattering function (IISF) $\phi_{\rho s}(q,t)$, and the incoherent longitudinal current correlation function (ILCCF) $\phi_{jls}(q,t)$.  These functions are discussed in paper I (see Eqs.\ (6)-(10)).  We are concerned with versions of these functions that are normalized to be unity at zero time.
 
 In the overdamped theory, the central theoretical function that must be evaluated is the irreducible memory kernel $m_{irr}$.  It has a graphical representation in terms of an infinite set of diagrams with a very restricted set of structures that is a consequence of taking the overdamped limit.  When an approximation for this function is obtained and combined with an approximation for the projected propagator in the generalized Enskog theory, the correlation functions for the fluid of interest can be calculated in a straightforward way.
 
 In Sec.\ \ref{sec:mirr}, we discuss the graphical representation of $m_{irr}$.  We focus on the simplest reasonable first approximation for  this function, which we call the one-loop approximation.
 In Sec.\ \ref{sec:kineticmodels}, we discuss the use of kinetic models to obtain an approximation for the generalized Enskog theory.  We focus on one of the simplest such models, which we call Model A. 
 In Sec.\ \ref{sec:stolA}, we discuss the calculation of the correlation functions of interest using the one-loop approximation and model A.
 In section \ref{sec:results}, the numerical results are presented for the Lennard-Jones fluid and compared to results from molecular dynamics simulations. Both a wide range of temperatures and wave vectors are considered at one high density. The overall strengths and weaknesses of the one-loop/Model A approximation are summarized in section \ref{sec:conclusions}, and some insight is provided into how the overdamped theory might be improved upon and extended to lower temperatures.

\section{Diagrammatic series for $m_{irr}$}\label{sec:mirr}

\subsection{The structure of diagrams in the series for $m_{irr}$}\label{sec:diagrammaticmirr}

The diagrammatic series for $m_{irr}$ is stated in Sec.\ VII F of paper I.  See Fig.\ 5 in paper I for examples of these diagrams.  Each diagram, except the simplest, contains one or more sets of vertices   with the following properties: (i) all the vertices in a set are singly connected to one another by $\chi_P^{ED}$ bonds, and  (ii) different sets are  connected by $\chi_{\hat0}^{(0)}$ bonds only.  All the points on  the vertices in a set will, in effect, have the same time associated with them because of the Dirac delta function time dependence of the $\chi_P^{ED}$ bonds connecting them.  Thus each set   can be viewed as an instantaneous vertex.  Since such a vertex contains two or more  of the fundamental vertices as well one or more $\chi_P^{ED}$ bonds, we shall refer to it as an instantaneous compound vertex. 

There are some vertices in a diagram that are not members of such a set. The vertex attached to the left root is always a $Q_{12}^{c1L}$ vertex that is not attached to any $\chi_P^{ED}$   bond in the diagram for $m_{irr}$.  The vertex attached to the right root may be a $Q_{21}^{c1L}$ vertex with no $\chi_P^{ED}$ bond attached, or it may be a  $T_{21}^{H}$ vertex that is part of a compound vertex.

 The general structure of   a diagram in the overdamped series for $m_{irr}$ can be described in the following way. 
 
 \noindent 1.\quad A diagram   consists of a collection of instantaneous vertices connected by $\chi_{\hat0}^{(0)}$ bonds. Each instantaneous vertex is either a single vertex or a compound vertex.
 
 \noindent 2.\quad The left root is attached to a $Q_{12}^{c1L}$ vertex. This vertex will be called the left endpoint vertex. 
 
 \noindent 3.\quad The right root is attached either to a $Q_{21}^{c1L}$ vertex or a compound vertex subject to certain restrictions. The vertex or compound vertex attached to the right root will be called the right endpoint vertex. 
 
 \noindent 4.\quad All other compound vertices are not attached to a root. They will be referred to as compound interaction vertices, and their structure is subject to certain restrictions. 

\noindent 5.  The topological restrictions  on the series for $m_{irr}$, stated in Sec. VII F of paper I, allow 

\noindent $(i)$ one type of left endpoint vertex, 

\noindent $(ii)$ several types of right endpoint vertices, and 

\noindent $(iii)$ many types of compound interaction vertices. 

\begin{figure}[h!]
\includegraphics[width=8.5cm,keepaspectratio=true]{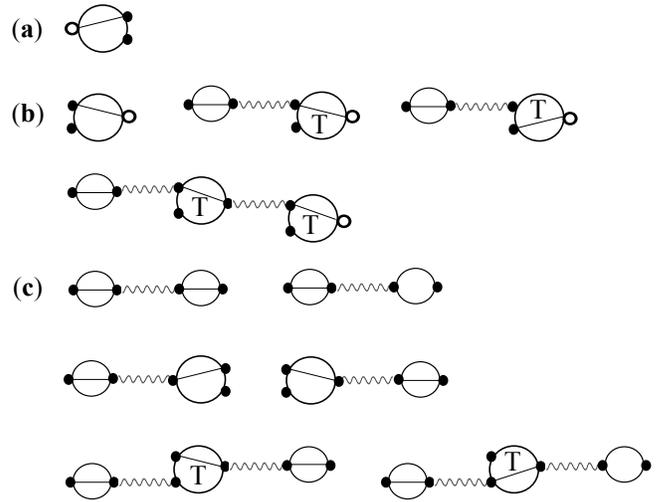}
\caption{Allowed endpoint vertices and allowed compound interaction vertices in the graphical series for $m_{irr}$.  (a)  The only allowed left endpoint vertex.  (b) Some of the simplest allowed right endpoint vertices.  (c) Some of the simplest allowed compound vertices.  Large open circles with the letter $T$ are $T_{21}^H$ vertices. Large open circles with no letter, one left point, and one right point are $Q_{11}^{c1}$ vertices  if they contain an internal line or $Q_{11}^{c0}$ vertices if they do not.  Large open circles with one left point, two right points, and an interior line are $Q_{12}^{c1L}$ vertices, and $Q_{21}^{c1L}$ vertices are represented similarly, but with the numbers of left and right points reversed.  Wavy lines are $\chi_P^{ED}$ bonds. The points are drawn as they would appear in an $m_{irr}$ diagram  with small open circles  for root points and small closed circles  for  free points. Free points that have no bond attached in this figure will have a $\chi_{\hat0}^{(0)}$ bond attached in the $m_{irr}$ diagram. In (b), the first vertex is a right endpoint vertex consisting of a $Q_{21}^{c1L}$ vertex with two left points and one right point. It is the only right endpoint vertex that is not a compound vertex. The second and third vertices contain a $T_{21}^H$ vertex and also have two left points and one right point. All other  right endpoint vertices are compound vertices with one right point and more than two left points. In (c), the first two vertices are the only two compound interaction vertices with only one left point and one right point.}
\label{fig:verticesinmirr}
\end{figure}

\noindent See Fig.\ \ref{fig:verticesinmirr} for examples of the structures of the allowed endpoint vertices and  compound interaction vertices. These considerations allow us to express the series for $m_{irr}$ in an alternative form. 
\vspace{1mm}

\noindent $m_{irr}(\mathbf{R},\lambda,t;\mathbf{R}^{\prime},\lambda^\prime,t^\prime)=$ the sum of all topologically distinct matrix diagrams with:

\noindent $(i)$ a left root labeled $(\mathbf{R},\lambda,t)$ and a right root labeled $(\mathbf{R}^{\prime},\lambda^\prime,t^\prime)$; 

\noindent $(ii)$ free points; 

\noindent $(iii)$ $\chi^{(0)}_{\hat0}$ bonds; 

\noindent $(iv)$ one allowed left endpoint vertex, one allowed right endpoint vertex, and allowed compound interaction vertices; 

\noindent such that:

\noindent $(i)$ the left root is attached to  the left endpoint vertex; 

\noindent $(ii)$ the right root is attached to the right endpoint vertex;

\noindent $(iii)$ each free point is attached to a bond and vertex;

\noindent $(iv)$ there is no bond whose removal would disconnect the roots.\,\,$\Box$
\vspace{0.5mm}

Although this function has two time arguments, its value depends only on the difference of the times.  In the discussion of the values of diagrams contributing to $m_{irr}$, we will set $t^\prime=0$ for convenience.

This diagrammatic series has an interesting feature. Since the only bonds are $\chi_{\hat0}^{(0)}$ bonds, when a diagram is evaluated, the Hermite index of every free point must be $\hat0$. (See Sec.\ IV of paper I for a discussion of the notation for Hermite indices.) In effect, the only dummy variable needed for a free point is  thus a position variable, meaning that the bonds and compound interaction vertices can be treated as functions of position only.   The endpoint vertices, on the other hand, still retain a         Hermite variable dependence, and consequently so does $m_{irr}$.
 
It follows from the discussion of the overdamped limit in paper I that every allowed left or right endpoint vertex is independent of $\nu$, every allowed compound interaction vertex is proportional to $\nu^{-1}$, and every compound interaction vertex contributes a factor of $t$ to the value of a diagram because of the integration over the time variable assigned to  it.   Thus the value of any diagram in the series above is of the form 
$$(t/\nu)^n\Theta(t)A(\mathbf{R},\lambda;\mathbf{R}^{\prime},\lambda^\prime),$$ 
where $n$ is the number of compound interaction vertices in the diagram.  Here $A$ is independent of $t$ and $\nu$ but is a function of the position and Hermite arguments of $m_{irr}$ as determined by the structure of the diagram.  $\Theta$ is the Heaviside function.  It follows that $m_{irr}$ has time dependence only on the long time scale of $O(\nu)$ that arises in the overdamped limit (see Sec.\ VII of paper I) and none on the shorter time scale of $O(\nu^{-1})$.  This makes the irreducible memory function very slowly varying for large $\nu$.

  Everything discussed in this subsection can be applied in a completely analogous fashion to the diagrams in the series for $m_{s\,irr}$, also defined in Sec. VIIF of paper I.  

\subsection{Evaluation of a compound interaction vertex}

Here we give an example of the evaluation of a compound interaction vertex.  The result for this specific example is important in the following development.

The first compound interaction vertex in (c) of Fig.\ \ref{fig:verticesinmirr} consists of two $Q_{11}^{c1}$ vertices connected by a $\chi_P^{ED}$ bond.  Imagine this as part of a diagram to be evaluated. Assign time arguments $t_1$ and $t_2$ to the left and right vertices, respectively, and position variables and Hermite indices to the points. The product of the functions associated with the vertices and bond are the following.
\begin{align*}
Q_{11}^{c1}({\bf R},\hat0;{\bf R}_1,\lambda_1)\chi_P^{ED}({\bf R}_1,\lambda_1,t_1;{\bf R}_2,\lambda_2,t_2)
\\*
\times Q_{11}^{c1}({\bf R}_2,\lambda_2;{\bf R}^\prime, \hat0)
\end{align*}
Here we have assigned $\hat0$ Hermite labels to the far left  and far right point because these points, in an $m_{irr}$ diagram, are always attached to a $\chi_{\hat0}^{(0)}$ bond, whose function is nonzero only for this Hermite index. Holding the $t_1$ time variable fixed, we then integrate over $t_2$, integrate over the position arguments of the $\chi_P^{ED}$, and sum over its Hermite arguments.  These are all steps that would be among those performed if a diagram in $m_{irr}$ containing this compound interaction vertex were being evaluated.   This gives the following.
\begin{align*}
&\int dt_2\,\sum_{\lambda_1.\lambda_2}\int d{\bf R}_1d{\bf R}_2\,Q_{11}^{c1}({\bf R},\hat0;{\bf R}_1,\lambda_1)
\\*
&\times\chi_P^{ED}({\bf R}_1,\lambda_1,t_1;{\bf R}_2,\lambda_2,t_2)Q_{11}^{c1}({\bf R}_2,\lambda_2;{\bf R}^\prime, \hat0)
\end{align*}
The $\chi_P^{ED}$ bond has a Dirac delta function time dependence.  See Sec.\ VII E of paper I.  The result of the time integration is
\begin{align}
&\sum_{\lambda_1.\lambda_2}\int d{\bf R}_1d{\bf R}_2\,Q_{11}^{c1}({\bf R},\hat0;{\bf R}_1,\lambda_1)
\nonumber
\\*
&\times\tilde\chi_P^{EO}({\bf R}_1,\lambda_1,0;{\bf R}_2,\lambda_2)Q_{11}^{c1}({\bf R}_2,\lambda_2;{\bf R}^\prime, \hat0)
\label{eq:firstcompoundinteractionvertex}
\end{align}
There is no time dependence in the above expression, and $\tilde\chi_P^{EO}$  is proportional to $\nu^{-1}$, so this compound interaction vertex is instantaneous and  proportional to $\nu^{-1}$  as noted above.
The second compound interaction vertex in (b) of Fig.\ \ref{fig:verticesinmirr} is very similar. Its value can be obtained from the expression in (\ref{eq:firstcompoundinteractionvertex}) simply by replacing the last $Q_{11}^{c1}$ by $Q_{11}^{c0}$.

These two compound interaction vertices are the only ones that have only one left point and one right point. The sum of the values of these compound interaction vertices is
\begin{align*}
&{\cal Q}_{11}({\bf R},{\bf R}^\prime)
 =\sum_{\lambda_1.\lambda_2}\int d{\bf R}_1d{\bf R}_2\,Q_{11}^{c1}({\bf R},\hat0;{\bf R}_1,\lambda_1)
\\*
&\times\tilde\chi_P^{EO}({\bf R}_1,\lambda_1,0;{\bf R}_2,\lambda_2)
\\*
&\quad\times\left(Q_{11}^{c1}({\bf R}_2,\lambda_2;{\bf R}^\prime, \hat0)+Q_{11}^{c0}({\bf R}_2,\lambda_2;{\bf R}^\prime, \hat0)\right)
\end{align*}
(A more explicit expression for this function requires information about $\tilde\chi_P^{EO}$. This is discussed in the next section.) This function, as defined, has no Hermite arguments. We  can define  a version  with Hermite arguments  as follows.
$${\cal Q}_{11}({\bf R},\lambda;{\bf R}^\prime,\lambda^\prime) ={\cal Q}_{11}({\bf R},{\bf R}^\prime)\delta_{\lambda\hat0}\delta_{\lambda^\prime\hat0}$$
It can be shown that $\hat{\cal Q}_{11}({\bf q})$ is a scalar that depends on the magnitude but not the direction of the wave vector ${\bf q}$.  

\subsection{Some symmetry properties of $m_{irr}$}

Other compound vertices can be evaluated in the same way.  Each compound interaction vertex is a scalar function of only the positions associated with its left and right points and is invariant with regard to rotation of the coordinate system. These positions are integrated over when a diagram that contains the compound vertex is evaluated.

A left or right interaction vertex is a function of the positions associated with its points.  It is also a function of the Hermite index associated with the root point to which it is attached.
As a result, each diagram in $m_{irr}$ is a function of the positions of its left and right roots and its Hermite indices and is equal to an integral over the positions associated with its free points. The only way in which the value of such a diagram can depend  on the orientation of the coordinate system is through the Hermite polynomial functions  used in calculating  its Hermite matrix elements. The general symmetry properties of the diagram values is rather complicated. Here we focus on the 3$\times 3$ array of matrix elements in which both the left and right Hermite indices are in the set ($\hat x,\hat y,\hat z$).  See Sec.\ IV of paper I for a discussion of Hermite matrix elements and indices.

This 3$\times$3 array of values obtained by assigning these Hermite indices to the roots of a specific diagram in $m_{irr}$ transforms as a second rank Cartesian tensor under rotation of the coordinate system. The Fourier transform of this array with regard to the position arguments on the roots is  a second rank Cartesian tensor that is a function of the wave vector. Moreover, it is straightforward to show that the elements of the array are invariant to those rotations of the coordinate system that do not rotate the wave vector. It follows that when the wave vector is in the $z$ direction, the 3$\times$3 array is diagonal and that the $(\hat x,\hat x)$ element is equal to the $(\hat y,\hat y)$ element. Thus $\hat{m}_{irr}(q\hat{\bf k},t)_{\lambda\lambda^\prime} =0 $
if $\lambda\neq\lambda^\prime$ and $\lambda,\lambda^\prime\in(\hat x,\hat y,\hat z)$. Also, $\hat{m}_{irr}(q\hat{\bf k},t)_{\hat x\hat x}=\hat{m}_{irr}(q\hat{\bf k},t)_{\hat y\hat y}
$.

\subsection{Renormalization of the propagator that appears in diagrams for $m_{irr}$}\label{sec:renormalization} 

 It is possible to show that if the fundamental correlation functions of interest decay to zero for long times (as is expected for an equilibrium system) the irreducible memory kernel must also decay to zero for long times.  As discussed above, each nonzero diagram in the series above for $m_{irr}$ has a value that is a nonnegative power of $t/\nu$.  Thus any approximation for $m_{irr}$ that includes only a finite number of diagrams in this series does not decay to zero for long positive times.  A series for $m_{irr}$ that can lead more easily to useful approximations can be obtained by performing a topological reduction that replaces the $\chi_{\hat0}^{(0)}$ propagator with a sum of chain diagrams containing $\chi_{\hat0}^{(0)}$ propagators and compound interaction vertices that have only one left and one right point.  To do this, we define the $\chi^O$ propagator in the following way.

\vspace{1mm}

\noindent$\chi^O({\bf R},t;{\bf R}^\prime,t^\prime)\equiv$ the sum of all topologically distinct matrix diagrams with:

\noindent $(i)$ a left root labeled $({\bf R},\hat0,t)$ and a right root labeled $({\bf R}^\prime,\hat0,t^\prime)$;

\noindent$(ii)$ free points;

\noindent$(iii)$ $\chi_{\hat0}^{(0)}$ bonds;

\noindent$(iv)$ ${\cal Q}_{11}$ vertices;

\noindent such that:

\noindent $(i)$ the left root is attached to a $\chi_{\hat0}^{(0)}$ bond;

\noindent $(ii)$ the right root is attached to a $\chi_{\hat0}^{(0)}$ bond;

\noindent $(iii)$ each free point is attached to a bond and a vertex.\quad$\Box$
\vspace{0.5mm}

\noindent The superscript $O$ is to denote that this propagator is defined in the overdamped limit. Note that this function as defined has no Hermite arguments, but the points in the graphs do have Hermite arguments. We also define a version of  this function with Hermite arguments.  
$$\chi^O({\mathbf R},\lambda,t;{\mathbf R}^\prime,\lambda^\prime,t^\prime)\equiv\chi^O({\mathbf R},t;{\mathbf R}^\prime,t^\prime)\delta_{\lambda\hat0}\delta_{\lambda^\prime\hat0}$$

We perform a topological reduction of the series for $m_{irr}$ that eliminates $\chi_{\hat0}^{(0)}$ bonds and replaces them with $\chi^{O}$ bonds.
\vspace{1mm}

\noindent $m_{irr}(\mathbf{R},\lambda,t;\mathbf{R}^{\prime},\lambda^\prime,t^\prime)=$ the sum of all topologically distinct matrix diagrams with:

\noindent $(i)$ a left root labeled $(\mathbf{R},\lambda,t)$ and a right root labeled $(\mathbf{R}^{\prime},\lambda^\prime,t^\prime)$; 

\noindent $(ii)$ free points; 

\noindent $(iii)$ $\chi^{O}$ bonds; 

\noindent $(iv)$ one allowed left endpoint vertex, one allowed right endpoint vertex, and allowed compound interaction vertices that have more than two points; 

\noindent such that:

\noindent $(i)$ the left root is attached to the left endpoint vertex; 

\noindent $(ii)$ the right root is attached to the right endpoint vertex;

\noindent $(iii)$ each free point is attached to a bond and vertex.\quad$\Box$
\vspace{0.5mm}

Each diagram in this series is a sum of an infinite number of diagrams in the previous series.  It follows from the discussion of the previous series that in the present series each diagram has a value of the form 
$$(t/\nu)^m\Theta(t)B(t/\nu;\mathbf{R},\lambda;\mathbf{R}^{\prime},\lambda^\prime).$$  Here $B$ is a power series in $t/\nu$ with nonnegative powers, and $m$ is the number of compound interaction vertices in the diagram.  The coefficients of the power series are functions of the position and Hermite arguments of $m_{irr}$.  The time dependence of $B$ reflects the time dependence of the $\chi^O$ propagator, which is causal and decays to zero as $t$ goes to infinity.  As a result, every diagram in this series goes to zero when $t$  goes to infinity, despite the positive powers of $t$ in the expression above.\cite{ftn:generalproof}  This makes it a more useful starting point for the construction of approximations for $m_{irr}$.

\subsection{The structure of diagrams in the series for $m_{irr}$ that has $\chi^O$ bonds}\label{sec:mirrstructure}

\begin{figure}[h!]
\includegraphics[width=8.5cm,keepaspectratio=true]{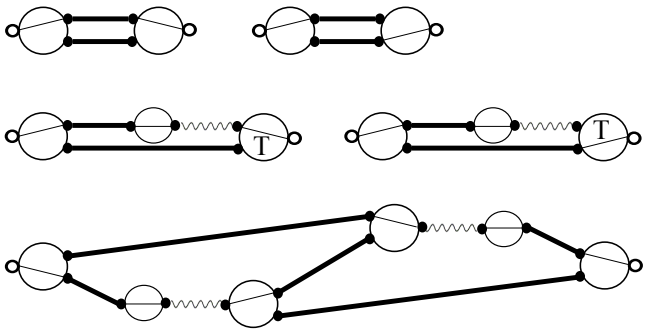}
\caption{Examples of diagrams in the series for $m_{irr}$ that has $\chi^{O}$ bonds. See Fig.\ \ref{fig:verticesinmirr} for the meaning of the various parts of a diagram. A  thick   solid line between two points on different vertices is a $\chi^O$ bond. The first two diagrams are one-loop diagrams that are nonzero at zero time. The next two diagrams are one-loop diagrams that are zero at $t=0$. The remaining diagrams in the series have compound interaction vertices and are also zero at $t=0$. The last diagram is an example of such a diagram.}
\label{fig:mirrdiagrams}
\end{figure}

  Figure \ref{fig:mirrdiagrams}  shows several  diagrams in the latest series for $m_{irr}$. 

The simplest diagrams (the first two in the figure) have a right endpoint vertex consisting of a $Q_{21}^{c1L}$ vertex and  just two $\chi^O$ bonds  with no compound interaction vertices. They can be described as one-loop diagrams.  There are just two such diagrams, and their values at $t=0$ are nonzero and depend on the range and magnitude of the longer ranged part of the potential of mean force.

The simplest diagrams that have a $T_{21}^H$ attached to the right root (the third and fourth diagrams in the figure) are also one-loop diagrams with only two $\chi^O$ bonds, but it can be shown that they are zero for $t=0$.  

Subsequent diagrams have one or more compound interaction vertices and three or more $\chi^O$ bonds. Each compound interaction vertex generates a power of $t/\nu$ in the value of the diagram for small $t$.  

For very large positive times $t$, each bond function decays to zero.  Therefore a large number of bonds  will likely cause the value of a diagram to decay rapidly at long times.  Thus it is plausible to expect that the most important diagrams for the longest times are those that have small numbers of compound interaction vertices.  It is also plausible to expect that the most important diagrams for the smallest times are those that are nonzero at $t=0$.

With this in mind, as a first approximation we keep only the diagrams that are nonzero for $t=0$ and that have the smallest number of bonds. They are the  two one-loop diagrams in Fig.\ \ref{fig:mirrdiagrams}  that have a $Q_{21}^{c1L}$ vertex  as their right endpoint vertex.    The value of these diagrams for $t\ge 0$ is of the form $B(t/\nu)$, where $B(x)$ is a power series whose leading nonzero term contains $x^0$.  It follows from the discussion in Sec.\ \ref{sec:renormalization} that the sum of the diagrams not included in this first approximation are of the form $C(t/\nu)$, where $C(x)$ is a power series whose leading nonzero term contains $x^1$.  Hence the diagrams retained in this first approximation should also  be the most important ones for the behavior of $m_{irr}$ for  $t$ of $O(\nu^0)$ or less. 

We shall refer to this approximation, for simplicity, as the one-loop approximation,  with the understanding that it contains only those one-loop diagrams that are nonzero at $t=0$.

\section{The use of a kinetic model}\label{sec:kineticmodels}

\subsection{The method of kinetic models}

Section VIA of Paper 1 gives the diagrammatic series for the Hermite matrix elements of $\chi_P^{E}$ and $\chi_{sP}^{E}$, the generalized Enskog projected propagator and projected self propagator. To carry out the calculation of the correlation functions of interest, these matrix elements are needed in the limit of large $\nu$. In principle, three steps would be involved in an exact calculation of these elements.

1.\  An exact calculation of all the matrix elements of $M^H$ and $M_s^H$.

2.\ The summation of all the graphs in the series for each of the elements of $\chi_P^{E}$ and $\chi_{sP}^{E}$.  These graphs contain $M^H$ and $M_s^H$ vertices.

3.\ Evaluation of the limiting behavior of these matrix elements for large $\nu$.  

Although the second and third steps are not problematic, the first step is extremely difficult because of the large number of matrix elements required.
In practice, the calculation must be done approximately.  

A useful method for constructing an approximation is the method of kinetic models,\cite{BHA54,GRO59,SUG68,MAZ72,FUR75,BoonYip} which in the context of the diagrammatic theory can be described in the following way.  

\noindent1.\quad Construct an approximation for the matrix elements of $M^H$ and $M_s^H$ that is consistent with the symmetry and dissipative properties of the exact matrices. The relevant symmetry properties are their behavior under rotation and inversion of the coordinate axes. The dissipative properties imply that all eigenvalues of the Fourier transforms of the two matrices are nonpositive. Such an approximation is called a kinetic model.

\noindent2.\quad Sum the series for the elements of $\chi_P^{ED}$ and $\chi_{sP}^{ED}$ containing the approximate versions of the $M^H$ and $M_s^H$ matrix elements.

The simplest of such models is the BGK model,\cite{BHA54} which in the present context can be constructed using the following assumptions.

\noindent1.\quad The $\hat{M}^H({\bf q})$ and $\hat{M}_s^H({\bf q})$ matrices are diagonal and independent of ${\bf q}$.

\noindent2.\quad The $\hat0\hat0$ matrix elements are zero. (See Sec.\ IV of paper I for a discussion of the notation for Hermite indices.)

\noindent3.\quad Every other diagonal element is equal to $-\nu$, where $\nu=-\hat{M}_{s}^H(\mathbf{q})_{\hat{z}\hat{z}}$.

Assumption 2 is correct. Also it is the case that the $\hat{M}_s^H$ matrix is independent of   $\mathbf{q}$.  The other components of these assumptions are simplifying approximations. 
The BGK model preserves the essential symmetry and dissipative properties of the exact $M^H$ functions, but it is too simple to describe the hydrodynamic behavior for small wave vector.

\subsection{Model A}
For the present calculation, we have constructed the simplest kinetic model that gives a reasonable description of the Enskog projected propagators for large wave vector and that has some of the hydrodynamic behavior of these propagators for small wave vector.  

\noindent 1.\quad All matrix elements of the exact $M^H$ and $M^H_s$ for which one or both indices is $\hat0$ are zero. (Here $\hat0=(000)$.)  We retain this feature in the kinetic model.  

\noindent 2.\quad Every nonzero element of  the exact $M^H$ and $M_s^H$ is $O(\nu)$, and we shall construct an approximation that retains this feature.

\noindent 3.\quad Consider the matrix elements of $M^H({\bf R};{\bf R}^\prime)$ and $M_s^H({\bf R};{\bf R}^\prime)$ among the basis functions $\hat x\equiv(100)$, $\hat y\equiv(010)$, and $\hat z\equiv(001)$.  It can be shown that, for both functions, these nine elements transform, under rotations of the coordinate system, as a symmetric second rank Cartesian tensor. Therefore, for each function, it is the sum of two terms: a scalar times the identity tensor and a symmetric traceless tensor. For $M_s^H$ the symmetric traceless part is exactly zero, and this is incorporated in the model. For $M^H$, we shall approximate the tensor of matrix elements by its first part. The symmetric traceless part contributes to the Enskog projected propagator, but to a significantly smaller extent.
 
\noindent 4.\quad All other diagonal matrix elements will be approximated as a negative constant of order $\nu$.

\noindent 5.\quad All other off diagonal elements will be set equal to zero.

\noindent 
We shall refer to this kinetic model as Model A.
 
 The matrix elements of $M^H$ and $M_s^H$ for Model A are the following. 
\begin{align*}
\lefteqn{M^H({\bf R},\lambda;{\bf R}^\prime,\lambda^\prime)}
\\*
&=0 
\quad\hbox{if $\lambda=\lambda^\prime=\hat0$ or $\lambda\neq\lambda^\prime$}
\\*
&=-\nu\left(\delta({\bf R}-{\bf R}^{\prime})-\frac{\delta(|{\bf R}-{\bf R}^\prime|-d)}{4\pi d^2}
\right)
\\*
&\quad\hbox{if $\lambda=\lambda^\prime=\hat x$ or $\hat y$ or $\hat z$}
\\*
&=-\nu_\lambda
\quad\hbox{if $\lambda=\lambda^\prime\ne\hat0$ or $\hat x$ or $\hat y$ or $\hat z$}
\\*
\lefteqn{M_s^H({\bf R},\lambda;{\bf R}^\prime,\lambda^\prime)}
\\*
&=0 
\quad\hbox{if $\lambda=\lambda^\prime=\hat0$ or $\lambda\neq\lambda^\prime$}
\\*
&=-\nu\delta({\bf R}-{\bf R}^{\prime})
\quad\hbox{if $\lambda=\lambda^\prime=\hat x$ or $\hat y$ or $\hat z$}
\\*
&=-\nu_\lambda
\quad\hbox{if $\lambda=\lambda^\prime\ne\hat0$ or $\hat x$ or $\hat y$ or $\hat z$}
\end{align*}
The corresponding Fourier transforms are
\begin{align*}
\lefteqn{\hat M^H({\bf q};\lambda,\lambda^\prime)}
\\*
&=0 
&&\hbox{if $\lambda=\lambda^\prime=\hat0$ or $\lambda\neq\lambda^\prime$}
\\*
&=-\nu\left(1-j_0(qd)
\right)
&&\hbox{if $\lambda=\lambda^\prime=\hat x$ or $\hat y$ or $\hat z$}
\\*
&=-\nu_\lambda
&&\hbox{if $\lambda=\lambda^\prime\ne\hat0$ or $\hat x$ or $\hat y$ or $\hat z$}
\\*
\lefteqn{\hat M_s^H({\bf q};\lambda,\lambda^\prime)}
\\*
&=0 
&&\hbox{if $\lambda=\lambda^\prime=\hat0$ or $\lambda\neq\lambda^\prime$}
\\*
&=-\nu
&&\hbox{if $\lambda=\lambda^\prime=\hat x$ or $\hat y$ or $\hat z$}
\\*
&=-\nu_\lambda
&&\hbox{if $\lambda=\lambda^\prime\ne\hat0$ or $\hat x$ or $\hat y$ or $\hat z$}
\end{align*}
Here $q=|{\bf q}|$, and $j_0$ denotes a spherical Bessel function. The quantity $\nu$ is the exact value of $-\hat{M}_s^H({\bf q})_{\hat z\hat z}$ (see Sec. VII A of paper I).  Each $\nu_\lambda$ is $O(\nu)$, but the precise values will not be needed for our calculation.
This simple kinetic model preserves many of the properties of the exact matrices.

\noindent 1.\quad  It preserves the rotational and translational symmetry of the exact memory functions.

\noindent2.\quad It preserves the fact that the eigenvalues of $\hat{M}^H({\bf q})$ and $\hat{M}_s^H({\bf q})$ are nonpositive for all ${\bf q}$.

\noindent3.\quad It preserves the fact that both functions have a zero eigenvalue for all ${\bf q}$ associated with the zero value of the $\hat0\hat0$ matrix element. This expresses the conservation of particles in a hard sphere collision. 

\noindent4.\quad It preserves the fact that three eigenvalues of $\hat{M}^H$ go to zero as ${\bf q}\to{\bf 0}$ by virtue of the fact that the three diagonal elements of $\hat{M}^H$ for $\lambda=\hat x$, $\hat y$, and $\hat z$ are nonzero for nonzero ${\bf q}$ and go to zero as ${\bf q}$ goes to zero.  This results from conservation of momentum in hard sphere collisions.  

\noindent5.\quad The matrix elements of $\hat{M}_s^H$ among the $\hat x$, $\hat y$, and $\hat z$ basis functions are exactly correct. The trace of these matrix elements for $\hat{M}^H$ is exactly correct.

The major qualitative deficiency of Model A is that it does not preserve the fact that there is a fourth eigenvalue of $\hat{M}^H$ that goes to zero as ${\bf q}\to{\bf 0}$. The corresponding eigenfunction is a linear combination of basis functions whose matrix indices $\lambda$ are (200), (020), and (002). This zero eigenvalue is a consequence of the conservation of kinetic energy in hard sphere collisions and is important for getting correct hydrodynamic behavior for small wave vector. This means, for example, that this model will be unable to capture the impact of sound waves on the density correlations of the liquid at small wave vectors, though it will at least describe correctly the fact that these fluctuations will decay slowly. For large wave vector, on the other hand, the model describes the predominantly dissipative behavior of the $\hat{M}^H$ function, and  it should be reasonably accurate for the $\hat{M}_s^H$ function at all wave vectors.

\subsection{Results for Model A}

The graphical series for $\chi_P^E$ in terms of $M^H$ is given in Sec.\ VI A of paper I.  
In the Fourier domain with Hermite matrix notation,  the diagrammatic series implies
$$\hat\chi_P^E({\bf q},t) =\exp\left([\hat Q_{11}^{c1}({\bf q})+\hat M^H({\bf q})]t\right)\hat\chi_P^{(0)}({\bf q},t)$$
$M^H(q)=O(\nu)$ and it is intrinsically negative. The only diagrams important in the overdamped limit are those that have no $Q_{11}^{c1}$ vertices, so for large $\nu$, the dominant behavior is
$$\hat\chi_P^{EO}({\bf q},t) =\exp\left(\hat M^H({\bf q})t\right)\hat\chi_P^{(0)}({\bf q},t)$$
$\hat M^H({\bf q})$ is a diagonal matrix for this kinetic model. $\hat\chi_P^{(0)}({\bf q},t)$ is  also diagonal, and its only time dependence is a Heaviside function. It follows that the diagonal elements of the dominant contribution to $\hat{\tilde\chi}_P^{EO}$ for Model A are:
\begin{align*}
\hat{\tilde\chi}_P^{EO}({\bf q},z)_{\lambda\lambda}
&=0 
&&\hbox{if $\lambda=\hat0$}
\\*
&=(z+\nu\left(1-j_0(qd)\right))^{-1}
&&\hbox{if $\lambda=\hat x$, $\hat y$, $\hat z$}
\\*
&=\left(z+\nu_\lambda\right)^{-1}
&&\hbox{if $\lambda\ne\hat0$, $\hat x$, $\hat y$, $\hat z$}
\end{align*}
\begin{align*}
\hat{\tilde\chi}_{sP}^{EO}({\bf q},z)_{\lambda\lambda}
&=0 
&&\hbox{if $\lambda=\hat0$}
\\*
&=(z+\nu)^{-1}
&&\hbox{if $\lambda=\hat x$, $\hat y$, $\hat z$}
\\*
&=(z+\nu_\lambda)^{-1}
&&\hbox{if $\lambda\ne\hat0$ , $\hat x$, $\hat y$, $\hat z$}
\end{align*}
The $z=0$ values are
\begin{align*}
\hat{\chi}_P^{ED}({\bf q})_{\lambda\lambda}
&=0 
&&\hbox{if $\lambda=\hat0$}
\\*
&=\left(\nu\left(1-j_0(qd)\right)\right)^{-1}
&&\hbox{if $\lambda=\hat x$, $\hat y$, $\hat z$}
\\*
&=\nu_\lambda^{-1}
&&\hbox{if $\lambda\ne\hat0$ or $\hat x$, $\hat y$, $\hat z$}
\\
\hat{\chi}_{sP}^{ED}({\bf q})_{\lambda\lambda}
&=0 
&&\hbox{if $\lambda=\hat0$}
\\*
&=\nu^{-1}
&&\hbox{if $\lambda=\hat x$, $\hat y$, $\hat z$}
\\*
&=\nu_\lambda^{-1}
&&\hbox{if $\lambda\ne\hat0$, $\hat x$, $\hat y$, $\hat z$}
\end{align*}
The off-diagonal elements are zero. (For the precise relationship between $\chi_P^{E}$, $\chi_P^{EO}$, and $\chi_P^{ED}$, see Sec.\ VII E of paper I.)

Using the above results, it can be shown that the ${\cal Q}_{11}$ vertex and $\chi^O$ bond have the following form when Model A is used.
\begin{align*}
\hat{\cal Q}_{11}({\bf q}) &=-\frac{v_T^2q^2}{\nu(1-j_0(qd))S(q)}
\\*
\hat{\cal Q}_{s11}({\bf q}) &=-\frac{v_T^2q^2}{\nu}
\\*
\hat{\chi}^O({\bf q},t) &=\Theta(t)\exp\left(-\left[\frac{v_T^2q^2}{\nu(1-j_0(qd))S(q)}\right]t\right)
\\*
\hat{\chi}_s^O({\bf q},t) &=\Theta(t)\exp\left(-\left[\frac{v_T^2q^2}{\nu}\right]t\right)
\end{align*}
In the above, $v_T$ is the thermal velocity $k_BT/m$, where $k_B$ is Boltzmann's constant, $T$ is the temperature, and $m$ is the mass of each particle in the liquid.  $S(q)$ is the static structure factor of the liquid, which can be calculated at any desired value of $q$ by numerical Fourier transform of $(g(r)-1)$.

\section{The  One-Loop Approximation and Model A}\label{sec:stolA}
\label{sec:ola}

\subsection{Evaluation of the diagrams in $m_{irr}$ that are included in the approximation}

\begin{figure}[h!]
\includegraphics[width=4cm,keepaspectratio=true]{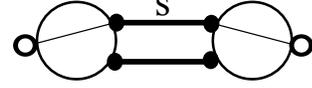}
\caption{The retained one-loop diagram for the self irreducible memory kernel.   See the captions of Figs.\ \ref{fig:verticesinmirr} and \ref{fig:mirrdiagrams} for the meanings of the various parts of the diagram.  Here, the bond between two vertices that has an S  is a $\chi_{s}^O$ bond.}
\label{fig:olsgraphs}
\end{figure}

\begin{figure}[h!]\includegraphics[width=8.5cm,keepaspectratio=true]{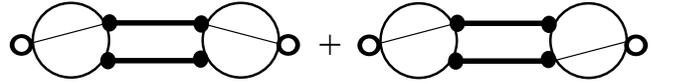}
\caption{The retained one-loop diagrams for the  irreducible memory kernel.  See the captions of Figs.\ \ref{fig:verticesinmirr} and \ref{fig:mirrdiagrams} for the meanings of the various parts of the diagram.}
\label{fig:oltgraphs}
\end{figure}

To compute any of the correlation functions of interest, we must first compute some approximation for the total and self irreducible memory kernels. To do this, we will evaluate only the simplest diagrams in their respective graphical series as discussed in Sec.\ \ref{sec:mirrstructure}. The diagram we retain for the self irreducible memory kernel is depicted in Figure \ref{fig:olsgraphs},  and the diagrams we retain for the total irreducible memory kernel are depicted in Figure \ref{fig:oltgraphs}.
 What these diagrams have in common is that they have one vertex on the left, one on the right, and two propagators connecting the  vertices that form a single internal loop. (There are no other vertices or propagators.) As noted above, we refer to this as the one-loop approximation.

The single diagram contributing to $m_{s\,irr}$ will be denoted as $m_{s\,1L}(\mathbf{R}_1,\lambda_1,t;\mathbf{R}_2,\lambda_2,0)$.  It has the following value.
\begin{align}
&m_{s1L}(\mathbf{R}_1,\lambda_1,t;\mathbf{R}_2,\lambda_2,0) \nonumber \\
&=\int d\mathbf{R}^{'}_{1}d\mathbf{R}^{'}_{2}d\mathbf{R}^{''}_{1}d\mathbf{R}^{''}_{2}\,
Q_{12}^{c1L}(\mathbf{R}_1, \lambda_1;\mathbf{R}^{'}_{1},\hat0,\mathbf{R}^{'}_{2},\hat0)
\nonumber
\\*
&\times
\chi_s^O(\mathbf{R}^{'}_{1},t;\mathbf{R}^{''}_{1})
\chi^O(\mathbf{R}^{'}_{2},t;\mathbf{R}^{''}_{2})
Q_{21}^{c1L}(\mathbf{R}^{''}_{1},\hat0,\mathbf{R}^{''}_{2},\hat0;\mathbf{R}_2,\lambda_2)
\end{align}
 Using the formulas for the $Q$ vertex matrix elements in Eqs.\ (C4) and (C5) of paper I, this can be expressed as the following, which is valid for $\hat i$ and $\hat j$ being $\hat x$, $\hat y$, or $\hat z$.
\begin{align} \label{eq:m1Lsdef}
&m_{s1L}(\mathbf{R}_1,\hat{i},t;\mathbf{R}_2,\hat{j},0) \nonumber \\
&=-\frac{\rho}{m}\chi_s^O(\mathbf{R}_1,t;\mathbf{R}_2)
\int d\mathbf{R}^{'}_{2}d\mathbf{R}^{''}_{2}\,\partial_{R_{1i}} v^{L}(\mathbf{R}_1-\mathbf{R}^{'}_{2})
\nonumber
\\*
&\times\chi^O(\mathbf{R}^{'}_{2},t;\mathbf{R}^{''}_{2}) 
\partial_{R^{''}_{2j}} e^{-v^{L}(\mathbf{R}^{''}_{2}-\mathbf{R}_2)/k_BT}
\end{align}
To calculate the correlation functions of interest, we need the $\hat{z}\hat{z}$ element of this function's Fourier transform. Using equation \eqref{eq:m1Lsdef} as a starting point, a complicated calculation yields the following result.
\begin{align} 
&\hat{m}_{s1L}(q\hat{\mathbf{k}},t)_{\hat{z}\hat{z}}=
\frac{4\pi\rho}{m}\int_{0}^{\infty}dR\,\chi_s^O(R,t) \nonumber \\
&\times\Biggl[G'(R,t)\left(\frac{2\cos(qR)}{q^2R}-
\frac{2\sin(qR)}{q^3R^2}\right)
\nonumber
\\*
&-G^{''}(R,t)\left(\frac{R\sin(qR)}{q}+
\frac{2\cos(qR)}{q^2}-\frac{2\sin(qR)}{q^3R}\right)\Biggr]
\label{eq:m1Lsresult}
\end{align}
The function $G(R,t)$ is a function whose Fourier transform is
\begin{equation} \label{eq:Gofqdef}
\hat G(q,t)=\hat{v}^{L}(q)\hat{\chi}^O(q,t)\mathcal{F}_{\mathbf{q}}
\left[e^{-v^{L}(R)/k_BT}-1\right]
\end{equation}
where $\mathcal{F}_{\mathbf{q}}$ is a functional that takes the Fourier transform of any function upon which it acts with respect to wave vector $\mathbf{q}$. The derivation of equation \eqref{eq:m1Lsresult} is detailed in the appendix.

The values of the two diagrams that contribute to $m_{irr}$ will be denoted $m_{1La}(\mathbf{R}_1,\lambda_1,t;\mathbf{R}_2,\lambda_2,0)$ and $m_{1Lb}(\mathbf{R}_1,\lambda_1,t;\mathbf{R}_2,\lambda_2,0)$, where $m_{1La}$ is the first diagram in Figure \ref{fig:oltgraphs} and $m_{1Lb}$ is the second. The value of $m_{1La}(\mathbf{R}_1,\lambda_1,t;\mathbf{R}_2,\lambda_2,0)$ is identical to equation \eqref{eq:m1Lsdef} with the first $\chi_s^O$ bond replaced by a $\chi^O$ bond. The value of $m_{1Lb}$ is 
\begin{align} 
&m_{1Lb}(\mathbf{R}_1,\hat{i},t;\mathbf{R}_2,\hat{j},0)
\nonumber
\\*
&=\frac{\rho}{m}
\int d\mathbf{R}^{'}_{2}d\mathbf{R}^{''}_{1},\left(\partial_{R_{1i}} v^{L}(\mathbf{R}_1-\mathbf{R}^{'}_{2})\right)\chi^O(\mathbf{R}_1,t;\mathbf{R}^{''}_{1})
\nonumber
\\*
&\times\chi^O(\mathbf{R}^{'}_{2},t;\mathbf{R}_2)
\partial_{R^{''}_{1j}} e^{-v^{L}(\mathbf{R}^{''}_{1}-\mathbf{R}_2)/k_BT}
\label{eq:m1Lbdef}
\end{align}
We need the $\hat{z}\hat{z}$ and $\hat{x}\hat{x}$ matrix elements of these two functions. We can get $\hat{m}_{1La}(q\hat{\mathbf{k}},t)_{\hat{z}\hat{z}}$ from equation \eqref{eq:m1Lsresult}, with the $\chi_s^O$ bond replaced by a $\chi^O$ bond. The function $\hat{m}_{1La}(q\hat{\mathbf{k}},t)_{\hat{x}\hat{x}}$ takes the form
\begin{align} 
&\hat{m}_{1La}(q\hat{\mathbf{k}},t)_{\hat{x}\hat{x}}
\nonumber
\\*
&=\frac{4\pi\rho}{m}\int_{0}^{\infty}dR\,\chi^O(R,t)\Biggl[G^{''}(R,t)\left(\frac{\cos(qR)}{q^2}-\frac{\sin(qR)}{q^3R}\right)
\nonumber
\\*
&+
G'(R,t)\left(\frac{\sin(qR)}{q^3R^2}-\frac{\cos(qR)}{q^2R}-\frac{\sin(qR)}{q}\right)\Biggr]
\label{eq:m1La100result}
\end{align}
The two matrix elements of $m_{1Lb}$ have the following expressions.
\begin{align} 
&\hat{m}_{1Lb}(q\hat{\mathbf{k}},t)_{\hat{z}\hat{z}}
\nonumber
\\*
&=\frac{4\pi\rho}{m}\int_{0}^{\infty}G^{'}_{1}(R,t)G^{'}_{2}(R,t)
\nonumber
\\*
&\times\left(\frac{R\sin(qR)}{q}+\frac{2\cos(qR)}{q^2}-\frac{2\sin(qR)}{q^3R}\right)
\label{eq:m1Lb001result}
\end{align}
\begin{align} 
&\hat{m}_{1Lb}(q\hat{\mathbf{k}},t)_{\hat{x}\hat{x}}
\nonumber
\\*
&=\frac{4\pi\rho}{m}\int_{0}^{\infty}dRG^{'}_{1}(R,t)G^{'}_{2}(R,t)
\left(\frac{\sin(qR)}{q^3R}-\frac{\cos(qR)}{q^2}\right)
\label{eq:m1Lb100result}
\end{align}
$G_1(R,t)$ and $G_2(R,t)$ are the inverse Fourier transforms of the following two functions
\begin{equation} \label{eq:G1ofqdef}
\hat{G}_1(q,t)=\hat{v}^{L}(q)\hat{\chi}^O(q,t)
\end{equation}
\begin{equation} \label{eq:G2ofqdef}
\hat{G}_2(q,t)=\mathcal{F}_{\mathbf{q}}\left[e^{-v^{L}(R)/k_BT}-1\right]\hat{\chi}^O(q,t)
\end{equation}
Once again, details on the derivation of these results may be found in the appendix.

\subsection{Numerical computation of correlation functions}\label{sec:numerical}

In this section we combine the one-loop approximation for $m_{irr}$ with Model A to describe how to calculate the correlation functions of interest for a monatomic fluid in the overdamped limit.

Appendix E of paper I gives the general equations for calculating correlation functions using the exact kinetic model for hard spheres.  These equations simplify greatly when Model A and the one-loop approximation are used.

In the one-loop approximation, the right endpoint vertex in all graphs that contribute to $m_{irr}$ is a $Q_{21}^{c1L}$ vertex.  The vertex must have left Hermite indices that are both $\hat0$.  The only Hermite matrix elements of $Q_{21}^{c1L}$ that satisfy this restriction are those whose right Hermite argument is either $\hat x$, $\hat y$, or $\hat z$.  (See Eq.\ (C5) of Paper I.)  The left endpoint vertex of all diagrams in $m_{irr}$ is a $Q_{12}^{c1L}$ vertex.  Similar reasoning applies to this vertex.  
Thus, in the one-loop approximation, $m_{irr\,\lambda\lambda^\prime}=0$ unless both Hermite indices are in the set ($\hat x$, $\hat y$, $\hat z$).  

More general arguments above show that the Fourier transform of this $3\times3$ array of functions is diagonal with its $\hat x\hat x$ element equal to its $\hat y\hat y$ element, provided the wave vector is in the $z$ direction.  

Thus it is convenient to make this choice for the wave vector, calculate the $\hat x\hat x$ and $\hat z\hat z$ elements of $\hat{m}_{irr}(q\hat{\bf k},t)$, and then use them to calculate the correlation functions. In fact, these matrix elements plus $\hat{\chi}_P^{ED}(q\hat{\bf k})$, which is obtained from the kinetic model, are all we need to calculate the correlation functions of interest.

In this subsection, we give the equations for doing this for the special case of the use of the one-loop approximation and Model A.

The irreducible memory function $M_{irr}$ is defined in Appendix E of paper I.  That definition plus the results discussed above show that $\hat{\tilde M}_{irr}(q{\bf k},z)$ is a $3\times3$ diagonal matrix.  The reducible memory function $M_{red}$ is also defined in Appendix E of paper I, and from that equation it is clear that $\hat{\tilde M}_{red}(q\hat{\bf k},z)$ is a $3\times3$ diagonal matrix.

\paragraph{Preliminary formulas.}  From above, for Model A, we have the following.
\begin{align}
&\hat{\cal Q}_{11}(q\hat{\bf k}) =-v_T^2q^2/\nu(1-j_0(qd))S(q)
\label{eq:ca}
\\*
&\hat{\chi}_P^{ED}(q\hat{\bf k})_{\lambda\lambda} =\left(\nu\left(1-j_0(qd)\right)\right)^{-1}\quad\hbox{for $\lambda=\hat z$ or $\hat x$.}
\label{eq:cb}
\\*
&\hat{\tilde\chi}_P^{EO}({\bf q},z)_{\lambda\lambda} =(z+\nu\left(1-j_0(qd)\right))^{-1}
\label{eq:cc}
\end{align}

\paragraph{Starting point for numerical calculation of correlation functions.}  Suppose we have calculated $\hat m_{irr}(q\hat{\bf k},t)_{\lambda\lambda}$ for various values of $q$ for $\lambda=\hat z$ and $\lambda=\hat x$ for the one-loop approximation. 
 
\paragraph{Calculation of $M^O$ and matrix elements of the projected propagator.}  The following equations are special cases of equations from Appendix E of paper I that take into account the simplifications to $m_{irr}$ that arise from the one-loop approximation and the kinetic model used.  They are written in the Fourier-Laplace domain for convenience, but they are converted to the Fourier-time domain for use in numerical calculations.  They hold for $\lambda=\hat z$ and $\lambda=\hat x$.
(Note that these are equations for specific matrix elements. There is no matrix multiplication involved.  Moreover, the calculations for different wave vectors or for different $\lambda$ are not coupled together.)  
\begin{align}
&\hat{\tilde M}_{irr}(q\hat{\bf k},z)_{\lambda\lambda} \equiv\hat{\tilde m}_{irr}(q\hat{\bf k},z)_{\lambda\lambda}\,\hat{\chi}^{EO}_{P}(q\hat{\bf k})_{\lambda\lambda}
\label{eq:c1}
\\*
&\hat{\tilde M}_{red}(q\hat{\bf k},z)_{\lambda\lambda} 
\nonumber
\\*
&=\hat{\tilde M}_{irr}(q\hat{\bf k},z)_{\lambda\lambda}+\hat{\tilde M}_{irr}(q\hat{\bf k},z)_{\lambda\lambda}\hat{\tilde M}_{red}(q\hat{\bf k},z)_{\lambda\lambda}
\label{eq:c2}
\\*
&\hat M^O(q\hat{\bf k},t) = {\cal Q}_{11}(q\hat{\bf k})\hat M_{red}(q\hat{\bf k},t)_{\hat z\hat z}
\label{eq:c3}
\\*
&\hat{\tilde\chi}_{P}(q\hat{\bf k},z)_{\lambda\lambda} =\hat{\chi}^{EO}_{P}(q\hat{\bf k})_{\lambda\lambda}+\hat{\chi}^{EO}_{P}(q\hat{\bf k})_{\lambda\lambda}\hat{\tilde M}_{red}(q\hat{\bf k},z)_{\lambda\lambda}
\label{eq:c4}
\end{align}
Eqs.\ (\ref{eq:c1})-(\ref{eq:c3}) are used to calculate numerical values of $\hat{M}_{irr}$, then $\hat M_{red}$, and then $\hat M^O$ for the various values of $q$ and for $\lambda=\hat z$ for a grid of times.  In these calculations $\hat\chi_P^{EO}$ in Eq.\ (\ref{eq:c1}) is replaced by $\hat{\chi}_P^{ED}(q\hat{\mathbf k})$ for simplicity.  Eqs.\ (\ref{eq:c1}), (\ref{eq:c2}), and (\ref{eq:c4}) are used to calculate numerical values of $\hat\chi_P(q\hat{\mathbf k},t)$ using $\lambda=\hat x$ and $\hat z$.  In these calculations, the equations were used as written, i.e.\ with $\hat\chi_P^{EO}$. 

\paragraph{Calculation of the propagator matrix elements associated with correlation functions of interest.}
In the overdamped limit, Eq.\ (11)  paper 1 becomes
\begin{align*}
&\frac{\partial \hat\chi(q\hat{\bf k},t)_{\hat0\hat0}}{\partial t}\nonumber
=\hat{\cal Q}_{11}(q\hat{\bf k})\hat\chi(q\hat{\bf k},t)_{\hat0\hat0}
\\*
&+
\int_0^t dt^{\prime}\,\hat M^O(q\hat{\bf k},t-t^\prime)\hat\chi({\bf q},t^{\prime})_{\hat0\hat0}
\quad\hbox{for $t>0$.}
\end{align*}
Numerical solution of this equation in the time domain gives results for the CISF.

Eq.\ (14) from paper 1 can be expressed in the following way.
\begin{eqnarray*}
\hat{\tilde\chi}(q\hat{\bf k},z)_{\lambda\lambda}&=&\hat{\tilde\chi}_P(q\hat{\bf k},z)_{\lambda\lambda}\left[1+\left(\hat{ Q}_{11}^{c1}(q\hat{\mathbf k})+\hat{Q}_{11}^{c0}(q\hat{\mathbf k})\right)_{\lambda\hat0}\right.
\nonumber
\\*
&&\times\left.\hat{\tilde\chi}(q\hat{\bf k},z)_{\hat0\hat0}\hat{Q}_{11}^{c1}(q\hat{\mathbf{k}})_{\hat0\lambda}
\hat{\tilde\chi}_P(q\hat{\bf k},z)_{\lambda\lambda}\right]
\nonumber
\\*
&&\quad\quad\hbox{for $\lambda=\hat z$ or $\hat x$}
\end{eqnarray*}
Evaluation of the right side of this equation in the time domain for the two values of $\lambda$ by numerical integration, using the numerical results for the ${\chi_P}_{\lambda\lambda}$  and $\chi_{\hat0\hat0}$ matrix elements obtained above, gives the LCCF and TCCF.

\paragraph{Self functions.}Every function used above in this section has a corresponding self function, with the exception of $Q$ vertices.  Every equation given above in this section has a `self' form that can be obtained by:

\noindent1.\ adding a subscript $s$ to the symbol for every function in the equation;

\noindent2.\ replacing each factor of $(1-j_0(qd))$ by 1;

\noindent3.\ replacing every factor of $S(q)$ by 1.

\noindent The self correlation function calculations require input data on the time dependence of $\hat m_{s\,irr}(q\hat{\bf k},t)$.

\section{Results}
\label{sec:results}

In this section, we test the theory based on the combination of the overdamped limit, the one-loop approximation, and Model A by comparing its results to molecular dynamics simulation data for an atomic liquid.\cite{tom1,tom2,ftn:MDTomYoung}
The liquid had a truncated and shifted Lennard-Jones potential of the form
\begin{align}
u(R)&=u_{LJ}(R)-u_{LJ}(R_c)&&R\leq R_c \nonumber \\
&=0, &&R>R_c
\end{align} 
where the cutoff distance $R_c$ is equal to $2.5\sigma$ and $u_{LJ}(R)$ is the standard Lennard-Jones potential given by
\begin{equation}
u_{LJ}(R)=4\epsilon\left(\left(\frac{R}{\sigma}\right)^{12}-
\left(\frac{R}{\sigma}\right)^6\right)
\end{equation}
For the duration of this paper, we will be using reduced Lennard-Jones units: $R^{*}=R/\sigma$, $t^{*}=(m\sigma^2/\epsilon)^{-1/2}t$, $\rho^{*}=N\sigma^3/V$, and $T^{*}=k_BT/\epsilon$.
Since we will be using these units exclusively, we will drop the asterisks in their definitions for ease of notation.

In order to compute the functions $v^{L}(R)$ and $S(q)$, we used molecular dynamics simulation data for the radial distribution function $g(R)$.\cite{ftn:grDas} These simulations used the velocity Verlet algorithm and were performed with $8000$ particles, a density of $0.85$, and temperatures of $0.723$, $1.554$, and $3.000$. Each simulation was one run, computing $g(R)$ up to a distance of $10$ with a resolution of $\delta r=0.01$. The selected density is that of the liquid near its triple point, and the temperatures range from the triple point to approximately twice the critical temperature.

In computing the contributions to the functions $\hat{m}_{irr}(q\hat{\mathbf{k}},\lambda,t;\lambda,0)$ and $\hat{m}_{s\,irr}(q\hat{\mathbf{k}},\lambda,t;\lambda,0)$, we took several numerical Fourier and inverse Fourier transforms, and all of these were performed using a discrete sine Fourier transform algorithm. The parameter $\nu$ was chosen to be the value that reproduces the time integral of the self RPMF memory function.\cite{ftn:RPMFJoyce} For more details, see Sec.\ VIII of paper I. The values for $\nu$ that we used are, from the
lowest temperature value to the highest, 9.82, 11.40, and 12.74.

The memory and self memory kernels were then used to compute the various correlation functions of interest.  To numerically solve the integro-differential equations given in section \ref{sec:numerical}, we used standard iterative methods, approximating the time integrals as left Riemann sums with a time interval of $\delta t=0.01$.

\begin{figure}[t]
\includegraphics[width=8.3cm,keepaspectratio=true]{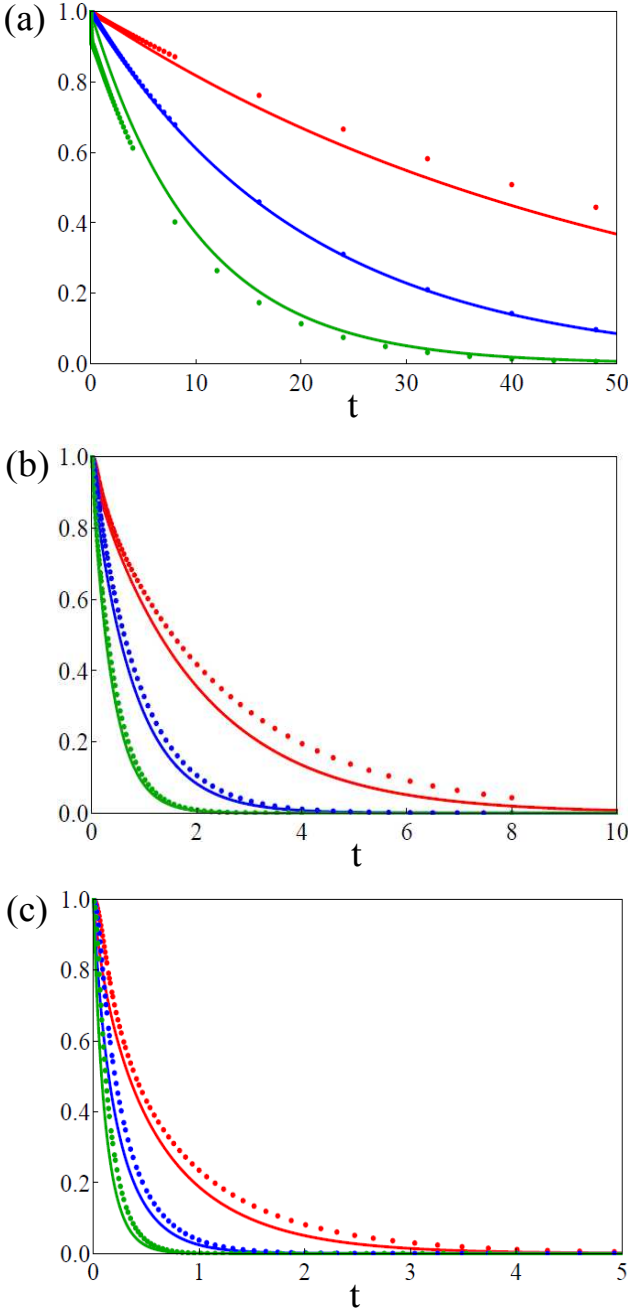}
\caption{The IISF as a function of time for fixed wave vector $q=0.75$ (graph a), $q=3.75$ (graph b), and $q=6.75$ (graph c). The points are the molecular dynamics simulation data, and the curves are computed from the approximate theory. In all three graphs, the top data set (red) is for $T=0.723$, the middle set (blue) is for $T=1.554$, and the bottom set (green) is for $T=3.000$.}
\label{fig:iisfgraphs}
\end{figure}
\begin{figure}[h!]
\includegraphics[width=8.5cm,keepaspectratio=true]{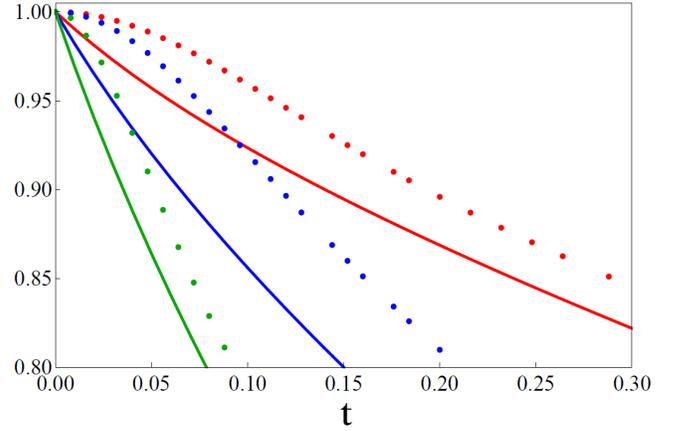}
\caption{The IISF as a function of time for fixed wave vector $q=3.75$ zoomed in to show the inaccuracy of the one-loop theory at very short times. The points are molecular simulation data, and the curves are the approximate theory. The top data set (red) is for $T=0.723$, the middle set (blue) is for $T=1.554$, and the bottom set (green) is for $T=3.000$.}
\label{fig:iisfgraphb}
\end{figure}
Figure \ref{fig:iisfgraphs} depicts the IISF as computed with the theory (solid curves) compared with the simulation data (points) at all three temperatures for wave vectors of $0.75$, $3.75$, and $6.75$. These three wave vectors represent the low, intermediate, and high wave vector regimes, respectively. The theory has very good quantitative agreement with the data, correctly describing the increasingly slow relaxation of the function as both temperature and wave vector are decreased. 

Figure \ref{fig:iisfgraphb} shows an expanded view of the middle graph of figure \ref{fig:iisfgraphs} for short times. The inaccuracy at short times is caused by the fact that the projected propagator at short times rapidly drops toward zero in a continuous fashion, whereas the calculations approximated this short time behavior as a Dirac delta function.  As a result, although the actual IISF has zero slope at $t=0$, the theory gives a negative slope.  On the scale of Fig.\ \ref{fig:iisfgraphs}, however, this very short time discrepancy is barely noticeable.

\begin{figure}[h!]
\includegraphics[width=8.3cm,keepaspectratio=true]{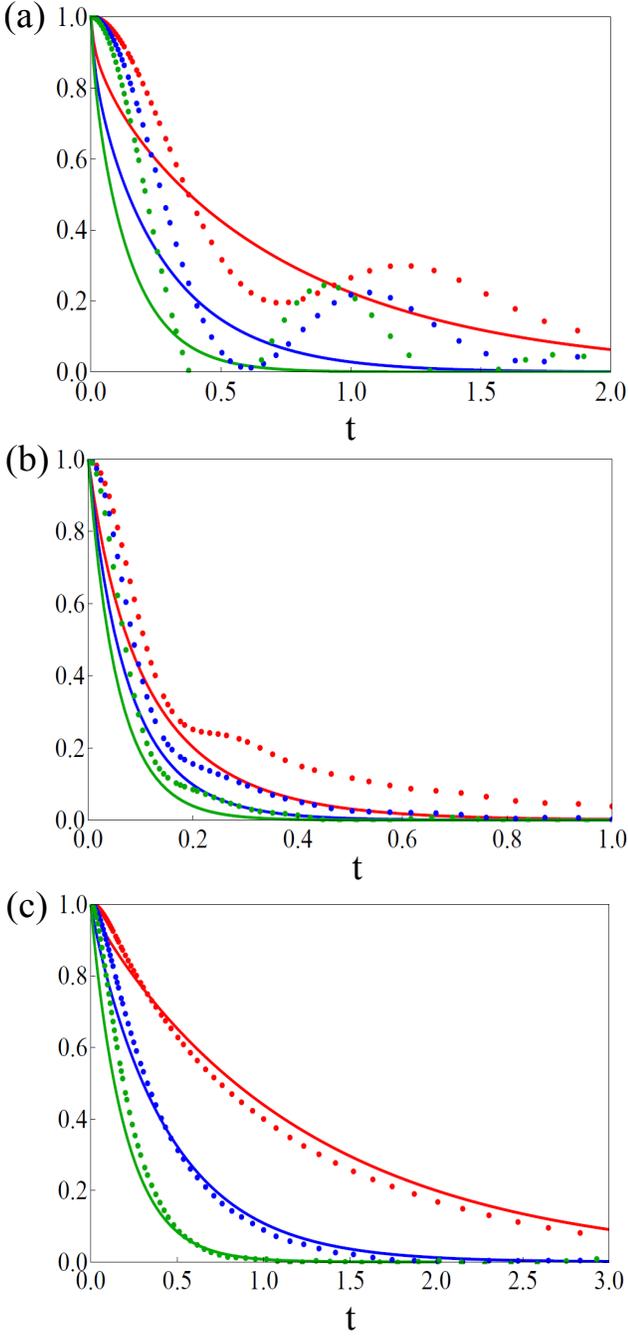}
\caption{The CISF as a function of time for fixed wave vector q=0.75 (graph a), q=3.75 (graph b), and q=6.75 (graph c). The points are the molecular dynamics simulation data, and the curves are computed from the approximate theory. In all three graphs, the top data set (red) is for T=0.723, the middle set (blue) is for T=1.554, and the bottom set (green) is for T=3.000.}
\label{fig:cisfgraphs}
\end{figure}

Figure \ref{fig:cisfgraphs} depicts the same comparison for the CISF. The theory does not correctly describe the CISF at small wave vectors, predicting a monotonic decay to zero, whereas the simulation results show oscillations that come from soundwave modes. Oscillatory hydrodynamic behavior of this sort is not accounted for by Model A. However, the time scale for relaxation at small wave vector is correctly given by the theory. As the wave vector increases and the impact of hydrodynamic effects is reduced, the theoretical results begin to improve, and the theory once again shows good quantitative agreement with the simulation results at a wave vector of $q=6.75$. 

\begin{figure}[h!]
\includegraphics[width=8.3cm,keepaspectratio=true]{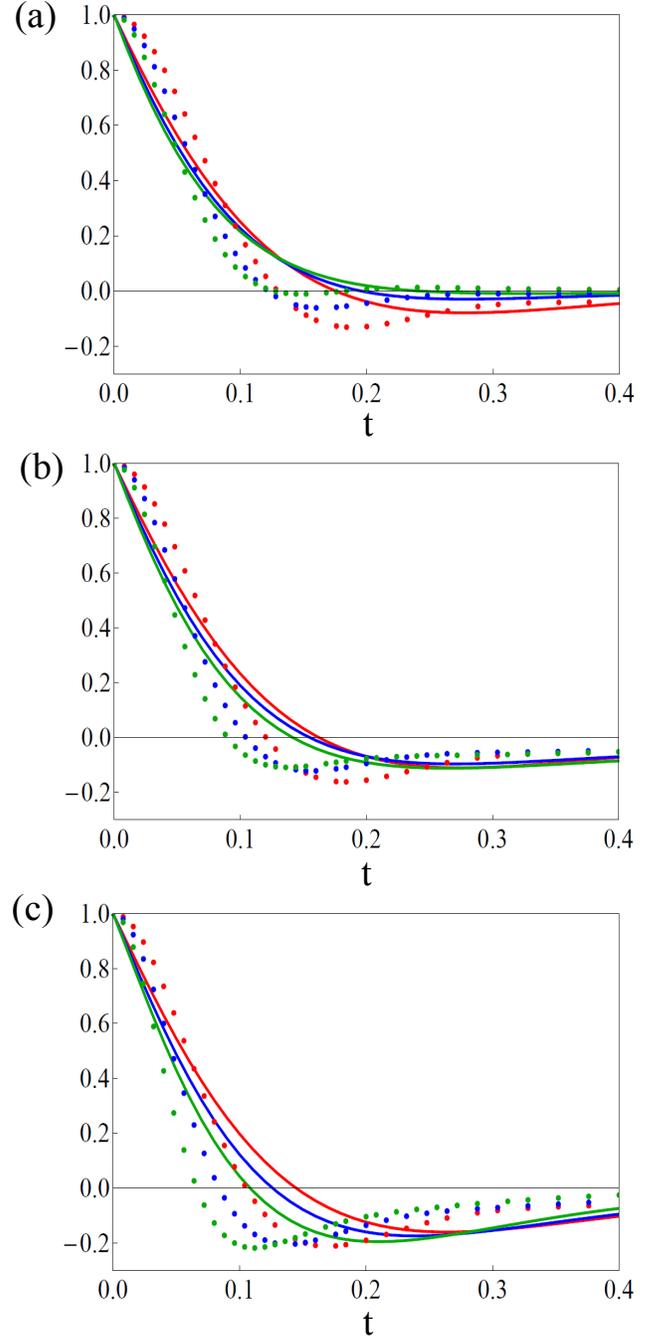}
\caption{The ILCCF as a function of time for fixed wave vector q=0.75 (graph a), q=3.75 (graph b), and q=6.75 (graph c). The points are the molecular dynamics simulation data, and the curves are computed from the approximate theory. In all three graphs, the top data set (red) is for T=0.723, the middle set (blue) is for T=1.554, and the bottom set (green) is for T=3.000.}
\label{fig:slccgraphs}
\end{figure}

Figure \ref{fig:slccgraphs} shows the same comparison for the ILCCF function. The theory correctly predicts slower relaxation as the temperature is lowered and as the wave vector is increased, but it still predicts the wrong slope at $t=0$. The  ILCCF decays to a value close to zero on a much shorter time scale than the IISF, however, crossing the horizontal axis by a time of around $0.1$, which is approximately equal to the value of $1/\nu$ for these states (see above).  The overdamped theory should not be expected to be accurate at such short times, and in fact it is not in good agreement with the data. For longer times, it is in fairly good agreement with the data, except for the longest wave vector.

\begin{figure}[h!]
\includegraphics[width=8.3cm,keepaspectratio=true]{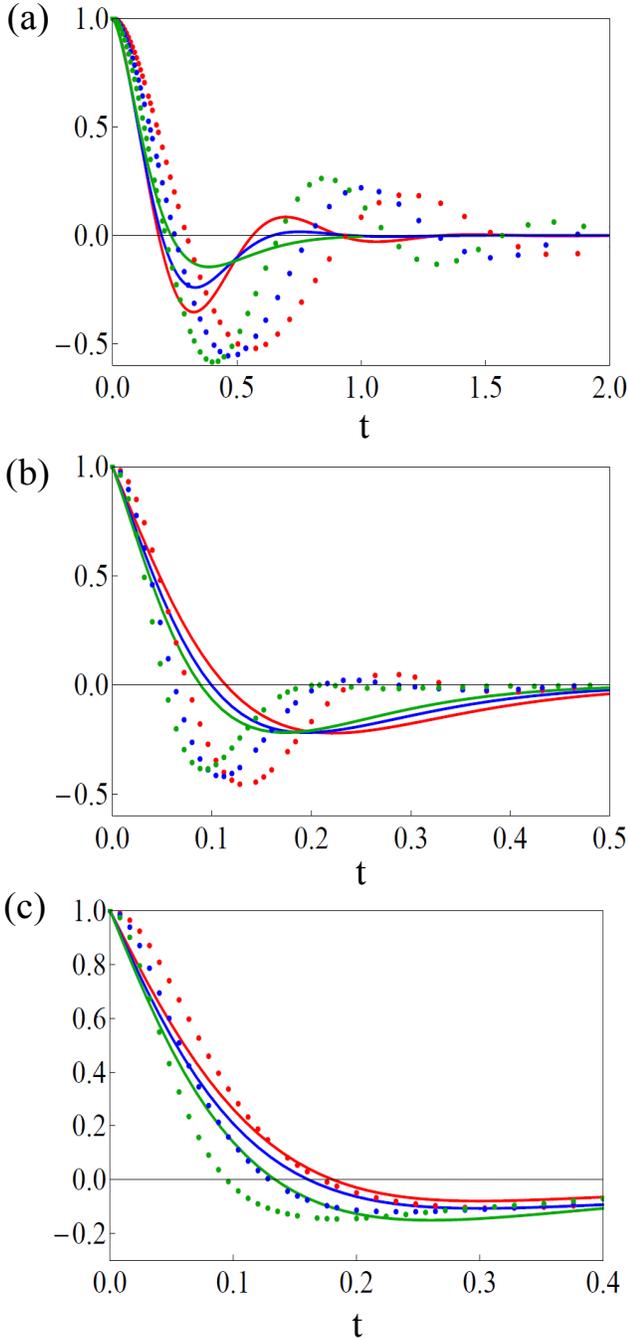}
\caption{The LCCF as a function of time for fixed wave vector q=0.75 (graph a), q=3.75 (graph b), and q=6.75 (graph c). The points are the molecular dynamics simulation data, and the curves are computed from the approximate theory. In all three graphs, the top data set (red) is for T=0.723, the middle set (blue) is for T=1.554, and the bottom set (green) is for T=3.000.}
\label{fig:lccgraphs}
\end{figure}

Figure \ref{fig:lccgraphs} shows the comparison for the LCCF function. As with the CISF, the theory does poorly at small wave vectors, failing to predict the slowly decaying oscillations in the simulation data.  These oscillations are once again due to hydrodynamic effects that are not accounted for by Model A. As the wave vector is increased, the theory becomes more accurate, at least qualitatively predicting the slower relaxation at lower temperatures. For longer times, it is in fairly good agreement with the data for the longest wave vector.

\begin{figure}[h!]
\includegraphics[width=8.5cm,keepaspectratio=true]{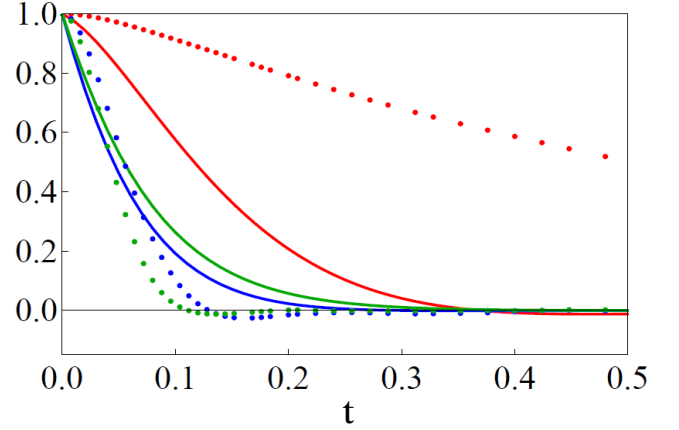}
\caption{The TCC as a function of time for fixed temperature T=3.000. The points are the molecular dynamics simulation data, and the curves are computed from the approximate theory. The red data set is for q=0.75, the blue set is for q=3.75, and the green set is for q=6.75.}
\label{fig:tccgraph}
\end{figure}

Figure \ref{fig:tccgraph} shows the comparison for the TCCF at a temperature of $3.000$ for the three representative wave vectors. The  theory predicts the slowest decay for wave vector $0.75$, but it severely underestimates the value of the function. The exceptionally slow decay at this wave vector is due to hydrodynamic-like shear modes, which once again are not described by Model A. The description of these modes may also require a more accurate approximation for the irreducible memory kernel than the one-loop approximation. As the wave vector is increased, the theory becomes more accurate, but it does not predict the small dip below the time axis exhibited by the data, and it incorrectly predicts the intermediate wave vector curve being below the high wave vector curve at times earlier than the crossover in the data at $t\approx 0.14$.
 
\section{Conclusions}
\label{sec:conclusions}

The approximate theory that is based on the overdamped kinetic theory, the one-loop approximation and Model A, correctly describes the temperature and wave vector dependence of many of the basic correlation functions of the Lennard-Jones liquid for the high density  and range of temperatures studied,
which extended from above the critical temperature down to the triple point temperature. This is especially the case for the incoherent intermediate scattering function at all wave vectors and the coherent intermediate scattering function at large wave vector. For the corresponding current correlation functions, the agreement is only qualitatively correct for the corresponding wave vectors and only at times longer than $1/\nu$, the average time for the
randomization of a particle's velocity by brief repulsive collisions.  The failure to describe the correlations functions well for times of $O(1/\nu)$ and less is to be expected, since the derivation of the theory makes it clear that the overdamped behavior takes place only for larger times.  The failure to describe the coherent functions (the CISF, LCCF, and the TCCF) for small wave vector is due to the fact that Model A is too simple a model of the hard sphere memory function to describe hydrodynamic oscillations correctly. The use of a more accurate model would presumably improve the results, though this would lead to a considerable increase in the complexity of the calculations required.

As temperature is lowered further, the CISF and the IISF will decay more and more slowly, and the one-loop approximation we used for the irreducible memory kernel will then probably become the weak link in the theory. Ideally, we would like to find ways of estimating the strengths of the vertices in the theory for temperatures well into the supercooled regime and calculate the behavior of the correlation functions for longer times and lower temperatures, but this will require the development of ways of finding the most important diagrammatic contributions to the series for the irreducible memory kernel at longer times.

The diagrammatic kinetic theory that is the basis of the current theoretical work describes a fluctuating equilibrium liquid in terms of the time and space dependent correlations of its density fluctuations. As the temperature is lowered into the supercooled regime, it is well recognized\cite{SIL99,GLO00,EDI00,RIC02,AND05} that dynamical heterogeneities become an important feature of the behavior of fluctuations and relaxation. The diagrammatic theory developed here and in the previous paper may provide a way of describing and understanding such dynamical heterogeneity in the case of atomic liquids.

\appendix
\section{Derivation of Major Results}

We start by considering the integral in Eq.\ \eqref{eq:m1Lsdef} for $\hat{i}=\hat{j}=001$. Let's call the integral that appears there $F(\mathbf{R},t)$, where $\mathbf{R}=\mathbf{R}_1-\mathbf{R}_2$. We can replace 
\begin{equation*}\exp(-v^{L}(\mathbf{R}^{\prime\prime}_{2}-\mathbf{R}_2)/k_BT)\end{equation*}
 with 
 \begin{equation}\exp(-v^{L}(\mathbf{R}^{\prime\prime}_{2}-\mathbf{R}_2)/k_BT)-1,
 \end{equation}
 since this function appears under a partial derivative. The reason for doing this is that the former function does not have a well-defined Fourier transform, since for large values of $\mathbf{R}^{\prime\prime}_{2}-\mathbf{R}_2$, it goes to one, rather than zero.

We can now take the Fourier transform of this integral. In performing this Fourier transform, we will take advantage of the fact that $v^{L}(\mathbf{R}^{''}_{2}-\mathbf{R}_2)$ depends only on the magnitude of its argument, i.e., it is equal to $v^{L}(\vert\mathbf{R}^{''}_{2}-\mathbf{R}_2\vert)=v^{L}(R)$. We do the Fourier transform with respect to the position vector $\mathbf{R}=\mathbf{R}_1-\mathbf{R}_2$, and without loss of generality we can set $\mathbf{R}_2$ to be zero, since our system is space translationally invariant. As such, the integral in equation \eqref{eq:m1Lsdef} takes on the form of a convolution of three functions, and the Fourier transform yields the following simple result.
\begin{equation} \label{eq:Fofqdef}
\hat F(\mathbf{q},t)=-q_{z}^{2}\hat{v}^{L}(q)\hat{\chi}^O(q,t)\mathcal{F}_{\mathbf{q}}
\left[e^{-v^{L}(R)/k_BT}-1\right]
\end{equation} 
In this expression, the functional $F_{\mathbf{q}}$ takes the Fourier transform of its argument with respect to a wave vector $\mathbf{q}$. Since $e^{-v^{L}(R)/k_BT}-1$ is a function of only the magnitude of $\mathbf{R}$, its Fourier transform will be a function of only the magnitude of $\mathbf{q}$. As such, the only orientation dependence of the above comes from the factor of $q_{z}^{2}$. Let us rewrite the above function in the following form.
\begin{equation}
\hat F(\mathbf{q},t)=-q_{z}^{2}\hat G(q,t)
\end{equation}

The function $\hat G(q,t)$ is defined in equation \eqref{eq:Gofqdef}. $G(R,t)$ is its inverse Fourier transform. 
\begin{equation}
G(R,t)=\int \frac{d\mathbf{q}}{(2\pi)^3} \hat G(q,t)e^{i\mathbf{q}\cdot\mathbf{R}}
\end{equation}
Taking two derivatives with respect to $R_z$ and using equation \eqref{eq:Fofqdef} allows us to relate $G(R,t)$ to $F(\mathbf{R},t)$.
\begin{equation}
\partial_{R_z}^{2}G(R,t)=-\int\frac{d\mathbf{q}}{(2\pi)^3}
q_{z}^{2}\hat G(q,t)e^{i\mathbf{q}\cdot\mathbf{R}}=F(\mathbf{R},t)
\end{equation}
If we use the fact that $\partial_{R_z}G(R,t)=(R_z/R)G'(R,t)$, where $G'(R,t)$ is $\partial G(R,t)/\partial R$, the above relation can be rewritten as
\begin{equation}
F(\mathbf{R},t)=\left(\frac{R_z}{R}\right)^2 G^{''}(R,t)+\frac{1}{R}G'(R,t)-
\frac{R_{z}^{2}}{R^3}G'(R,t)
\end{equation}
Plugging this result back into Eq.\ \eqref{eq:m1Lsdef}, we find that
\begin{align} 
&m_{s1L}(\mathbf{R},t)_{\hat{z}\hat{z}}
=\frac{\rho}{m}\chi_{s}^{O}(R,t)\nonumber
\\*
&\times\left[\frac{R_{z}^{2}}{R^3}G'(R,t)-
\left(\frac{R_z}{R}\right)^2 G^{''}(R,t)-\frac{1}{R}G'(R,t)\right].
\label{eq:m1LsRzGform}
\end{align}

Now we can perform the Fourier transform with $\mathbf{q}\parallel\hat{\mathbf{k}}$. With this choice of the orientation of $\mathbf{q}$, $\mathbf{q}\cdot\mathbf{R}$ becomes $qR\cos\theta$, where $\theta$ is the angle between $\mathbf{R}$ and the $z$-axis. The appearance of the polar angle $\theta$ suggests that we perform the three dimensional Fourier transform integral in spherical coordinates, so using the fact that $R_z=R\cos\theta$ and substituting $\mu=\cos\theta$, we get
\begin{align*}
&\hat m_{s1L}(q\hat{\mathbf{k}},t)_{\hat{z}\hat{z}}
=\frac{\rho}{m}\int_{0}^{2\pi}d\phi\int_{-1}^{1}d\mu
\int_{0}^{\infty}dR
\nonumber
\\*
&\times\left[\mu^2R G'(R,t)-\mu^2R^2G^{''}(R,t)-RG'(R,t)\right] 
\nonumber
\\*
&\times\chi_{s}^{O}(R,t)e^{-iqR\mu}
\end{align*}
The angular integrals can be performed rather easily, leaving us with the result quoted in Eq.\ \eqref{eq:m1Lsresult}.

We now move on to Eq.\ \eqref{eq:m1Lbdef}, again for $\hat{i}=\hat{j}=\hat{z}$.  This expression can be rewritten as the product of two integrals.
\begin{align} \label{eq:m1LbF1F2}
&m_{1Lb}(\mathbf{R}_1,t;\mathbf{R}_2)_{\hat{z}\hat{z}}
\nonumber
\\*
&=\frac{\rho}{m}\int d\mathbf{R}^{\prime}_{2}
\partial_{R_{1z}} v^{L}(\mathbf{R}_1-\mathbf{R}^{\prime}_{2})\chi^O(\mathbf{R}^{\prime}_{2},t;\mathbf{R}_2) \nonumber \\
&\times\int d\mathbf{R}^{\prime}_{1} \partial_{R^{''}_{1z}} e^{-v^{L}(\mathbf{R}^{''}_{1}-\mathbf{R}_2)/k_BT}
\chi^O(\mathbf{R}_1,t;\mathbf{R}^{''}_{1}) \nonumber \\
&=\frac{\rho}{m}F_1(\mathbf{R},t)F_2(\mathbf{R},t)
\end{align}
where
\begin{equation}
F_1(\mathbf{R},t)=\int d\mathbf{R}^{\prime}_{2}
\partial_{R_{1z}} v^{L}(\mathbf{R}_1-\mathbf{R}^{\prime}_{2})\chi^O(\mathbf{R}^{\prime}_{2},t;\mathbf{R}_2)
\end{equation}
and
\begin{equation}
F_2(\mathbf{R},t)=\int d\mathbf{R}^{\prime\prime}_{1} \partial_{R^{\prime\prime}_{1z}} e^{-v^{L}(\mathbf{R}^{''}_{1}-\mathbf{R}_2)/k_BT}
\chi^O(\mathbf{R}_1,t;\mathbf{R}^{\prime\prime}_{1})
\end{equation}
Following the same procedure as before, we can take the Fourier transforms of these functions to get the following results.
\begin{equation}
\hat F_1(\mathbf{q},t)=iq_z \hat v^{L}(q)\hat\chi^O(q,t)=iq_z\hat G_1(q,t)
\end{equation}
\begin{equation}
\hat F_2(\mathbf{q},t)=iq_z\mathcal{F}_{\mathbf{q}}\left[e^{-v^{L}(R)/k_BT}-1\right]\hat\chi^O(q,t)
=iq_z\hat G_2(q,t)
\end{equation}
The functions $\hat G_1(q,t)$ and $\hat G_2(q,t)$ can be inverse Fourier transformed to yield $G_1(R,t)$ and $G_2(R,t)$.
These two functions are related to $F_1(\mathbf{R},t)$ and $F_2(\mathbf{R},t)$ by the following expressions.
\begin{equation}
F_1(\mathbf{R},t)=\frac{R_z}{R}G^{'}_{1}(R,t)
\end{equation}
\begin{equation}
F_2(\mathbf{R},t)=\frac{R_z}{R}G^{'}_{2}(R,t)
\end{equation}
Plugging these two results back into equation \eqref{eq:m1LbF1F2}, we get
\begin{equation} \label{eq:m1LbG1G2}
m_{1Lb}(\mathbf{R},t)_{\hat{z}\hat{z}}=\frac{\rho}{m}\left(\frac{R_z}{R}\right)^2
G^{'}_{1}(R,t)G^{'}_{2}(R,t)
\end{equation}
Now we take the Fourier transform of the above with $\mathbf{q}\parallel\hat{\mathbf{k}}$. We get
\begin{align*}
&\hat{m}_{1Lb}(q\hat{\mathbf{k}},t)_{\hat{z}\hat{z}}
=\frac{\rho}{m}\int_{0}^{2\pi}d\phi\int_{-1}^{1}d\mu
\int_{0}^{\infty}dR
\\*
&\times\mu^2R^2G^{'}_{1}(R,t)G^{'}_{2}(R,t)e^{-iqR\mu}
\end{align*}
The angular integrals can once again be performed analytically, and the end result is Eq.\ \eqref{eq:m1Lb001result}.

The functions $m_{1La}(\mathbf{R},t)_{\hat{x}\hat{x}}$ and $m_{1Lb}(\mathbf{R},t)_{\hat{x}\hat{x}}$ can be obtained from Eqs.\ \eqref{eq:m1LsRzGform} and \eqref{eq:m1LbG1G2} respectively if one replaces $R_z$ with $R_x$. Taking the Fourier transforms of these expressions with $\mathbf{q}\parallel\hat{\mathbf{k}}$, using 
the relation $R_x=R\sin\theta\cos\phi$, and performing the angular integrals gives equations \eqref{eq:m1La100result} and \eqref{eq:m1Lb100result}.


\begin{thebibliography}{10}
\bibitem{ftn:generalproof} A general proof that $\chi^O$ decays to zero when $t$ goes to infinity requires a detailed discussion of the Enskog theory for hard spheres.  In the next section, an explicit approximation for this propagator is obtained from an approximation to Enskog theory, and this approximation has that property.
\bibitem{ftn:MDTomYoung} The molecular dynamics calculations were performed by Tom Young.
\end{thebibliography}
\end{document}